\documentclass[runningheads]{llncs}
\usepackage{graphicx}
\usepackage{subfigure}
\usepackage{amsfonts}
\usepackage[colorlinks]{hyperref}
\usepackage{marvosym}
\usepackage{floatrow}
\usepackage{amsmath} 
\usepackage{multirow}
\newfloatcommand{capbtabbox}{table}[][0.48\textwidth]
\begin{document}
%
% \title{Unsupervised Rigid Registration of DSA/CTA Aortic Images}
\title{UDCR: Unsupervised Aortic DSA/CTA Rigid Registration Using Deep Reinforcement Learning and Overlap Degree Calculation}

\titlerunning{Unsupervised Aortic DSA/CTA Rigid Registration}

\author{Wentao Liu\inst{1}\and Bowen Liang\inst{2} \and Weijin Xu\inst{1} \and Tong Tian\inst{3}\and  Qingsheng Lu\inst{2} \and  Xipeng Pan\inst{4} \and  Haoyuan Li\inst{1} \and  Siyu Tian\inst{5} \and Huihua Yang \inst{1,4(\textrm{\Letter})} \and Ruisheng Su \inst{6} }
\authorrunning{W. Liu et al.}

\institute{School of Artificial Intelligence, Beijing University of Posts and Telecommunications, Beijing, China\\
\email{liuwentao@bupt.edu.cn}\\
\and Department of Vascular Surgery, Shanghai Changhai Hospital, Naval Medical University, Shanghai, China \\
\and Laboratory of Structural Analysis, Optimization and CAE Software for Industrial Equipment, School of Aeronautics and Astronautics, Dalian University of Technology, Dalian University of Technology, Dalian, China\\
\and School of Computer Science and Information Security, Guilin University of Electronic Technology, Guilin 541004, China
\and Ultrasonic Department, The Fourth Hospital of Hebei Medical University and Hebei Tumor Hospital, Shijiazhuang, China
% \url{http://www.springer.com/gp/computer-science/lncs}
\and Department of Radiology \& Nuclear Medicine, Erasmus MC, University Medical Center Rotterdam, The Netherlands
}
%%%%%%%%%%%%%Anonymous%%%%%%%%

% \author{Anonymous}
% \authorrunning{Anonymous}
% \institute{Anonymous Organization\\
% \email{**@******.***}
% }

%
\maketitle           % typeset the header of the contribution
\begin{abstract}

The rigid registration of aortic Digital Subtraction Angiography (DSA) and Computed Tomography Angiography (CTA) can provide 3D anatomical details of the vasculature for the interventional surgical treatment of conditions such as aortic dissection and aortic aneurysms, holding significant value for clinical research. However, the current methods for 2D/3D image registration are dependent on manual annotations or synthetic data, as well as the extraction of landmarks, which is not suitable for cross-modal registration of aortic DSA/CTA. In this paper, we propose an unsupervised method, UDCR, for aortic DSA/CTA rigid registration based on deep reinforcement learning. Leveraging the imaging principles and characteristics of DSA and CTA, we have constructed a cross-dimensional registration environment based on spatial transformations. Specifically, we propose an overlap degree calculation reward function that measures the intensity difference between the foreground and background, aimed at assessing the accuracy of registration between segmentation maps and DSA images. This method is highly flexible, allowing for the loading of pre-trained models to perform registration directly or to seek the optimal spatial transformation parameters through online learning. We manually annotated 61 pairs of aortic DSA/CTA for algorithm evaluation. The results indicate that the proposed UDCR achieved a Mean Absolute Error (MAE) of 2.85 mm in translation and 4.35° in rotation, showing significant potential for clinical applications.

\end{abstract}
\section{Introduction}
% In clinical practice, surgeons employ Digital Subtraction Angiography (DSA) as a visual guide to maneuver guidewires, catheters, and other interventional tools during the execution of interventional procedures~\cite{imaging}. 
% This surgery is characterized by rapid recuperation, pronounced therapeutic outcomes, and a diminished risk of complications,
Cardiovascular diseases are the leading killer jeopardizing global human health and are also one of the significant factors contributing to human disability~\cite{global1}. Endovascular intervention, a minimally invasive procedure that utilizes real-time imaging guidance to precisely deliver therapeutic devices through the vascular system to target and treat vascular pathologies, is widely applied in the treatment of vascular diseases involving the aorta, coronary, and intracranial arteries. Nevertheless, 2D Digital Subtraction Angiography (DSA) imaging guidance struggle to furnish 3D morphological information of vascular structures~\cite{su2022cave,dias}. This deficiency could impede physicians' comprehensive grasp of the guidewire's trajectory and the pathological areas, thus undermining the precision and efficacy of surgical procedures. The 2D/3D registration has the capability to achieve spatial fusion between preoperative Computed Tomography Angiography (CTA) 3D images and real-time 2D intra-interventional DSA images, holds promising potential for addressing this issue~\cite{markelj2012review}.

% The process of registering these images with 2D DSA, to achieve the spatial fusion of 3D pre-interventional images and real-time 2D intra-interventional images, holds promising potential for addressing this issue.
% Despite the limitations posed by the 2D imaging of intraoperative procedures, Computed Tomography Angiography (CTA) and Magnetic Resonance Angiography (MRA), which are non-invasive imaging techniques used for the preoperative diagnosis of CVDs, offer 3D anatomical details of the vasculature~\cite{imaging}. 
The registration of 2D/3D images has been explored for several decades, with comprehensive techniques having been developed for specific clinical application~\cite{svecic2021multimodal}. The traditional 2D/3D registration methods commonly formulate the registration task as an optimization problem, which utilize feature or intensity-based metrics to identify the overlap between Digitally Reconstructed Radiography (DRR) of 3D surface meshes or volumes and the target 2D images~\cite{review}. However, such methods face limitations in DSA/CTA registration. The divergent imaging principles of DSA and CTA present challenges in establishing a precise similarity metric, leading to a propensity for solutions to settle into local optima. Recently, deep learning based regression methods have been proposed for the 2D/3D registration, which leverage Convolutional Neural Network (CNN) to learn the 6DoF pose of the X-ray source~\cite{guan2020transfer}. Unfortunately, they rely on paired and annotated X-ray and volumetric images for training. Given the challenge of acquiring clinical annotated data, it is common practice to employ techniques such as DRR to synthesize X-ray images for the training of regression models~\cite{huang2024real}. Nevertheless, the efficacy of these models in registering real clinical data often does not achieve satisfactory results. The vascular centerline, as a key marker of vascular morphology information, has been utilized as a target for 2D/3D registration in several methodologies~\cite{zhu2019monte,zhu2021iterative,CAR-Net},  However, the extraction of the central line is contingent upon the results of DSA segmentation. For the aorta, angiography in DSA often exhibits incomplete visualization, complicating the extension of centerline-based registration methods to aortic applications.

% particularly in the realm of coronary arteries~\cite{}.necessitating the prior construction of corresponding CTA and DSA vascular segmentation datasets. 

To address these issues, we propose an unsupervised aortic DSA/CTA rigid registration method, UDCR, using Deep Reinforcement Learning (DRL), which is independent of ground truth or synthetic data and does not necessitate segmentation or centerline extraction of DSA. Specifically, we have established a cross-dimensional 2D/3D DRL registration environment based on spatial transformations, predicated on the imaging principles of DSA. Importantly, we propose a novel overlap degree calculation reward function designed to evaluate the registration accuracy between segmentation maps and DSA images. This function quantitatively assesses the degree of overlap for the foreground (vessel) and the background, respectively, which provides a metric for the precision of registration. Experimental results demonstrate that UDCR is capable of both pre-training and online learning, yielding impressive registration results.

\section{Method}
\begin{figure}[t]
\centering
\includegraphics[scale=0.45]{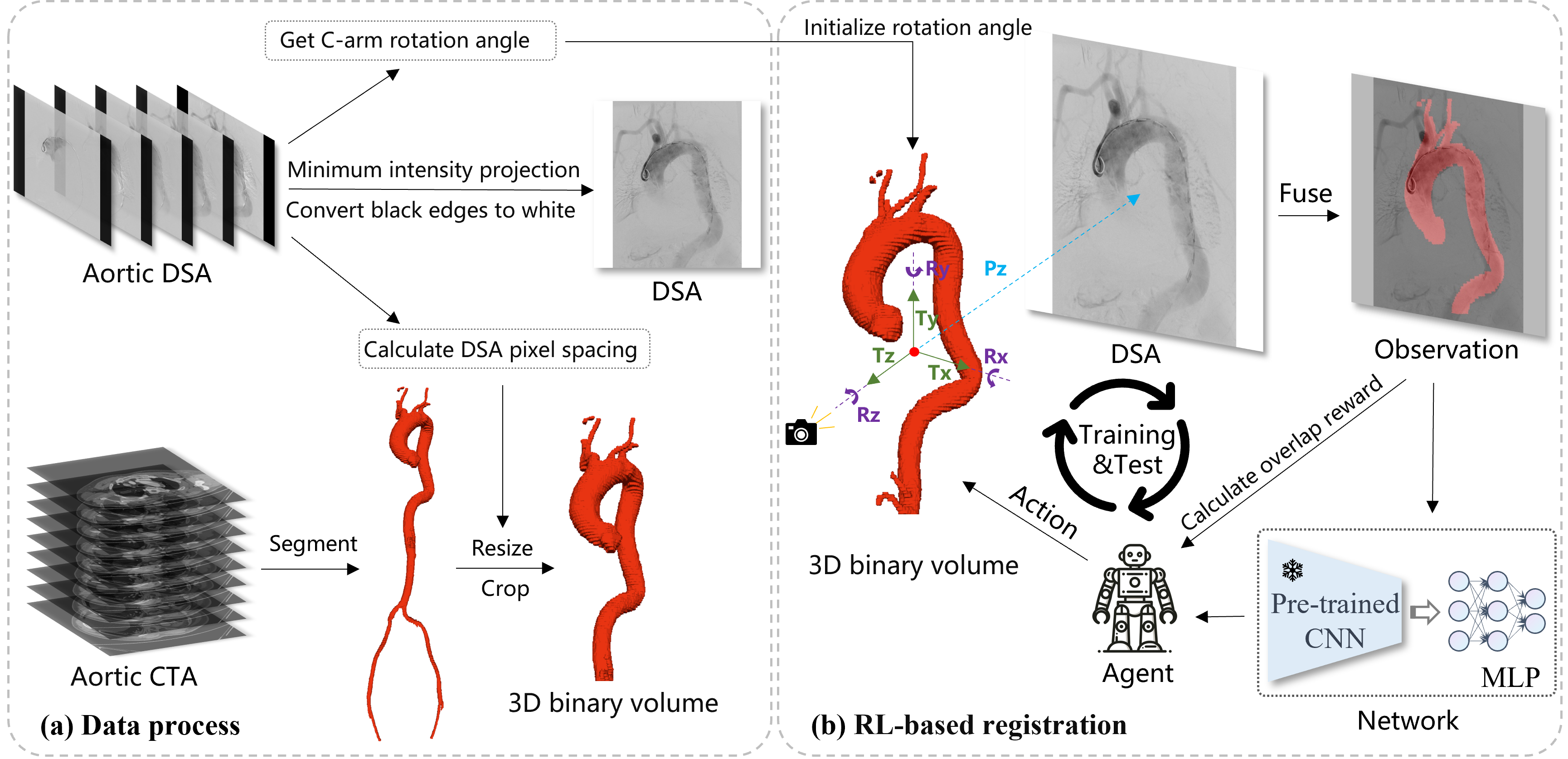}
\caption{Deep reinforcement learning-based rigid registration of aortic DSA/CTA} \label{fig:method}
\end{figure}
\subsection{Problem Definition}

% In 2D/3D registration, the goal is to find an optimal spatial transformation $Topt$ of the volume V from the observed X-ray projections $Irf$ such that when the images are overlaid, there is minimal misalignment~\cite{jaganathan2023self}. The problem can be formulated as an optimization problem with an objective function F that minimizes the misalignment as described:

% This problem is formulated as an    problem, aiming to minimize misalignment through an objective function \(F\).

% The goal is to minimize this function to achieve as close a match as possible between the transformed volume and the X-ray projection.

In 2D/3D registration, the objective is to identify an optimal spatial transformation $\mathbf{T_{opt}}$ that adjusts the volume $\mathbf{V} $ based on observed X-ray projections $\mathbf{I}_f $, ensuring minimal misalignment between the overlaid images~\cite{jaganathan2023self}. The mathematical formulation is:

\begin{equation}
\underset{T}{\operatorname{argmin}} \; \mathcal{F}(\mathbf{I}_f, \mathcal{S}(\mathbf{V}, \mathbf{T})),
\end{equation}
where $\mathcal{F}$ is the objective function to measure the misalignment between two images.
$\mathcal{S}(\mathbf{V}, \mathbf{T})$ is a function representing the application of transformation $\mathbf{T}$ to volume $\mathbf{V}$.

In the aortic DSA/CTA rigid registration, $\mathbf{I}_f$ represents the intraoperative aortic DSA, while $\mathbf{V}$ corresponds to the aortic CTA acquired preoperatively. During interventional procedures, the operator manipulates the C-arm by translating and rotating it to capture the patient's aortic DSA from the optimal viewing angle, facilitating the conduct of interventional surgical treatment. Consequently, the spatial transformation operation $\mathcal{S}$ includes 3D rotation and translation in three directions, as well as projection.

\subsection{Data Preprocessing} 
\label{Preprocessing}
Before registration, it is necessary to preprocess the paired aortic DSA and CTA images. We begin by reading the pixel array from the DICOM files of the DSA. A Minimum Intensity Projection (MinIP) is applied by taking the minimum value of each pixel across the tamporal dimension in DSA. This process serves to preserve the contrast agent information to the greatest extent, thereby converting the DSA sequence into a single 2D DSA image that highlights the lumen of the vessels by depicting the areas with the least intensity, which usually correspond to the contrast-filled regions. This preprocessing step is crucial for enhancing the visibility of the vascular structures and facilitating the subsequent registration with the CTA images. Considering the calculation method of the registration reward function described in section~\ref{DSA/CTA Registration Environment}, we convert the black edges in the MinIP to white. Moreover, the actual pixel spacing $S_{\text{DSA}}$ of the DSA images is calculated by multiplying the 
\textit{pixel spacing} with an \textit{estimated radiographic magnification factor} from the DSA. 

To extract aorta from CTA images, we employ a hybrid architecture, PHTrans~\cite{phtrans}, under the nn-UNet~\cite{nnunet} framework for the training of aortic segmentation on the AVT dataset~\cite{avt}. The trained model is subsequently applied to segment the aorta in CTA images prepared for registration. Furthermore, the aortic segmentation volume is resampled using nearest neighbor interpolation to adjust its sizes. This step ensures that the spatial resolution of the segmented volume matches the actual pixel spacing of the DSA, laying the foundation for precise registration. The sizes of resampling are calculated based on $S_{\text{CTA}} \times D_{\text{CTA}}/S_{\text{DSA}}$, where $S_{\text{CTA}}$ and $ D_{\text{CTA}}$ is the original pixel spacing and dimensions of the CTA. Additionally, rotation angle information extracted from the DSA is employed to initialize the rotational transformation of CTA before registration. This preparatory step is essential for setting up an accurate starting point for the rotational alignment of the DSA and CTA, facilitating a more efficient and precise registration process.

\subsection{DSA/CTA Registration Environment} 
\label{DSA/CTA Registration Environment}
Using DRL for DSA/CTA registration involves training a DRL agent to find the optimal spatial alignment between 2D DSA and 3D CTA. The agent iteratively adjusts the spatial transformation of the CTA to maximize alignment with the DSA image, guided by a reward function that quantifies the quality of the registration. This process leverages state representations that include transformation parameters and potentially image-derived features, exploring a defined action space of rotations and translations. The objective is to accurately overlay vascular structures from DSA onto the detailed anatomical context provided by CTA, with the learning algorithm refining the agent's policy based on feedback to converge on the most effective registration policy.

% enhance the precision of vascular interventions by
% \subsubsection{Action and Observations}

% \vspace{7pt} 
% \noindent\textbf{Action and Observations}
\subsubsection{Action and Observations}
The spatial transformation for 2D/3D registration involves a 6DoF  pose adjustment and scaling. Yet, in clinical practice, the rotation of the C-arm used for aortic DSA is confined to the $Z$ and $Y$ axes, as shown in Fig.~\ref{fig:method} (b). The translation along the $T_z$ direction is utilized to adjust the distance between the patient and the X-ray source, thereby controlling the spacing of DSA imaging. In our registration implementation, the binary segmented volume from the CTA has already been resampled during preprocessing to match the spacing of the DSA. Furthermore, the conversion from 3D to 2D employs parallel projection, meaning the spacing of the DSA remains unchanged. Therefore, the translation along $T_z$ is omitted. Consequently, the final registration actions include rotations $R_z$, $R_y$, translations $T_y$, $T_x$, and parallel projection.  The action space for the agent at each step is defined as $[[-T_x, T_x], [-T_y, T_y], [-R_z, R_z], [-R_y, R_y]]$. Given a binary segmentation volume 
$\mathbf{V}$, the spatial transformation is as follows:
\begin{equation}
\mathbf{P} = \min_{z}  (\mathbf{R}  (\mathbf{T} \mathbf{V}))
\end{equation}
where, the matrix \(\mathbf{T}\) denotes translation transformations defined by the displacements $[T_x, T_y]$, while $\mathbf{R}$ represents a rotation matrix, which consists of rotations around the $z-$ and $y-$axes as indicated by $R_z$ and $ R_y$.
$\min_{z}$ denotes the operation of taking the minimum value along the z-axis direction, which achieves the MinIP of volume $\mathbf{V}$ onto the image $\mathbf{P}$.   

The 3D binary volume undergoes a spatial transformation to become a 2D binary image, which is subsequently centered and aligned with the DSA image. Based on this, we have devised two types of observations: 1) Concat: perform a matrix concatenation of the binary image with the DSA. 2) Fuse: convert the binary image into a color image and proceed to fuse it with the DSA image, ensuring the preservation of the image's original dimensions.

% \begin{equation}
% \mathbf{R}_y(\theta_y) = \begin{bmatrix} \cos\theta_y & 0 & \sin\theta_y \\ 0 & 1 & 0 \\ -\sin\theta_y & 0 & \cos\theta_y \end{bmatrix}
% \end{equation}

% \begin{equation}
% \mathbf{R}_z(\theta_z) = \begin{bmatrix} \cos\theta_z & -\sin\theta_z & 0 \\ \sin\theta_z & \cos\theta_z & 0 \\ 0 & 0 & 1 \end{bmatrix}
% \end{equation}

% \begin{equation}
% \mathbf{R} = \mathbf{R}_z(\theta_z) \cdot \mathbf{R}_y(\theta_y) \cdot \mathbf{R}_x(\theta_x)
% \end{equation}

% \begin{equation}
% \mathbf{T} = \begin{bmatrix} 1 & 0 & 0 & T_x \\ 0 & 1 & 0 & T_y \\ 0 & 0 & 1 & 0 \\ 0 & 0 & 0 & 1 \end{bmatrix}
% \end{equation}

% \subsubsection{Reward Fuction}

% \vspace{7pt} 
% \noindent\textbf{Reward Fuction}
\subsubsection{Reward Fuction}
The registration reward function is crucial for guiding the DRL agent towards the optimal spatial alignment between DSA and CTA, which quantitatively evaluates the quality of registration at each step, providing feedback that is instrumental in refining the agent's policy for spatial transformation. In DSA, the contrast agent delineating the vasculature appears as black, serving as the foreground, while the background, processed through digital subtraction, appears as white. Building upon this imaging characteristic, we propose a reward function to calculate the degree of overlap for the foreground (vessel) and the background, respectively, providing a metric for the precision of registration, which is articulated as follows:

\begin{equation}
R = - \log \left( \frac{\sum_{i,j : P_{ij} \times F_{ij} > 0} P_{ij} \times F_{ij}}{\sum_{i,j : P_{ij} \times F_{ij} > 0} P_{ij}} \Bigg/ \frac{\sum_{i,j : (1 - P_{ij}) \times F_{ij} > 0} (1 - P_{ij}) \times F_{ij}}{\sum_{i,j : (1 - P_{ij}) \times F_{ij} > 0} (1 - P_{ij})} \right)
\end{equation}
% \begin{equation}
% \text{R} = - \log \left( \frac{\sum\limits_{i,j:P_{ij} \times F_{ij} > 0} P_{ij} \times F_{ij}}{\sum\limits_{i,j} P_{ij} \times F_{ij} > 0} \Bigg/ \frac{\sum\limits_{i,j:(1-P_{ij}) \times F_{ij} > 0} (1-P_{ij}) \times F_{ij}}{\sum\limits_{i,j} (1-P_{ij}) \times F_{ij} > 0} \right)
% \end{equation}
where $P$ is a binary vascular map with values of 0 and 1, and $F$ represents the grayscale image of the DSA. The product $P_{ij} \times F_{ij}$ is an operation that zeroes out the DSA pixel intensities at the locations designated as background in the binary map. Consequently, the first term of the formula calculates the mean intensity of DSA pixels situated at the foreground locations of the binary map. Similarly, the second term computes the mean intensity of DSA pixels at the background positions of the binary map. Theoretically, the greater the registration accuracy between the DSA and the binary map, the lower the value of the first term and the higher the value of the second term. By dividing these two values and taking the negative logarithm, it ensures a positive correlation between reward value and registration performance.

\section{Experiments}
\subsection{Dataset}
The data for the registration of aortic DSA/CTA originated from *** Hospital, comprising endovascular interventional DSA images for aortic aneurysms and aortic dissection, as well as preoperative and postoperative CTA examinations, totaling 61 paired cases. We developed a manual registration interactive program, enabling vascular surgeons to perform manual registration on all data preprocessed as described in section~\ref{Preprocessing}. The spatial transformation parameters obtained through this process serve as the ground truth for assessing the accuracy of the registration method. We select 40 pairs of data as the training set and the remaining 21 pairs as the test set.

\subsection{Implementation Details}

We performed aortic DSA/CTA registration experiments utilizing Proximal Policy Optimization (PPO) and Soft Actor-Critic (SAC), which represent state-of-the-art on-policy and off-policy reinforcement learning algorithms, respectively. The experiments were conducted using PyTorch on a single GeForce RTX A100 GPU, and for the implementation of the SAC and PPO, we employed stable baselines~\cite{stable-baselines3}. In the experiments, the learning rate was set to 3e-4, the batch size to 64, and the total number of timesteps to 10000. The maximum translation and rotation angle for the agent at each step are 5 pixels and 0.5$^\circ$, respectively. We employed two types of policy networks: one is the CnnPolicy from the stable baselines library, and the other (PCM) combines a pre-trained CNN with a Multilayer Perceptron (MLP). During the training phase, the pre-trained CNN component is frozen, allowing only the MLP part to be updated. This approach leverages the powerful representational capabilities of CNN for visual tasks while focusing on the optimization of policy decision-making through the MLP.

The metric for evaluating registration accuracy included reward values and the Mean Absolute Error (MAE) between the spatial transformation parameters relative to the starting point and the manually set ground truth. Considering the advantage of not relying on ground truth, the UDCR registration method exhibits considerable flexibility. We can employ UDCR for pre-training on the training dataset, where the refined policy is utilized for real-time registration. Moreover, UDCR can be directly employed for online learning. We conducted experiments on these two modes, configuring them through different approaches including observation, DRL algorithms, and policy networks. The spatial transformation parameters that yield the maximum reward are selected for evaluation.

\begin{table}[tbp]
\centering
\caption{ Results of the comparative experiment. (PT: PreTrain, OL: Online Learning)}\label{tab:vs}
\renewcommand\arraystretch{1.3}
\tabcolsep2.9pt
\scriptsize

\begin{tabular}{ccccccccccc}
\hline
\multirow{3}{*}{Mode} & \multicolumn{5}{c}{Methods}                                                                & \multirow{3}{*}{Reward $\uparrow$} & \multicolumn{4}{c}{\multirow{2}{*}{MAE $\downarrow$}} \\ \cline{2-6}
                      & \multirow{2}{*}{DRL} & \multicolumn{2}{c}{Observations} & \multicolumn{2}{c}{Networks} &                         & \multicolumn{4}{c}{}                         \\ \cline{3-6} \cline{8-11} 
                      &                     & Concat          & Fuse          & CNN              & PCM             &               & $T_y$ (mm)    & $T_x$ (mm)    & $R_z$ ($^\circ$) & $R_y$ ($^\circ$)   \\ \hline
\multirow{6}{*}{PT}   & SAC                 & \checkmark      &               & \checkmark       &                 & .244$\pm$.109 & 46.6$\pm$29.9 & 21.7$\pm$16.5 & 4.3$\pm$3.5 & 8.6$\pm$4.7  \\
                      & SAC                 &                 &\checkmark     & \checkmark       &                 & .242$\pm$.108 & 46.6$\pm$29.9 & 21.7$\pm$16.5 & 4.0$\pm$3.3 & 6.9$\pm$4.5  \\
                      & SAC                 &                 & \checkmark    &                  & \checkmark      & .342$\pm$.115 & 12.6$\pm$18.5 & 16.4$\pm$13.6 & 5.4$\pm$4.1 & 6.8$\pm$5.7 \\ \cline{2-11} 
                      & PPO                 & \checkmark      &               & \checkmark       &                 & .362$\pm$.089 & 13.9$\pm$11.7 & 8.8$\pm$9.0  & 3.2$\pm$2.6  & 8.6$\pm$7.5\\
                      & PPO                 &                 &\checkmark     &\checkmark        &                 & .334$\pm$.111 & 23.7$\pm$23.1 & 12.5$\pm$12.2 & 6.7$\pm$5.3  & 10.1$\pm$9.0 \\
                      & PPO                 &                 & \checkmark    &                  & \checkmark      & .364$\pm$.096 & 9.4$\pm$7.0   & 14.4$\pm$13.1 & 4.6$\pm$4.2  & 8.8$\pm$6.8  \\ \hline
\multirow{6}{*}{OL}   & SAC                 &\checkmark       &               & \checkmark       &                 & .445$\pm$.095 & 1.7$\pm$1.0   & 4.3$\pm$3.5   & 3.0$\pm$3.1  & 6.0$\pm$4.0  \\
                      & SAC                 &                 & \checkmark    & \checkmark       &                 & .444$\pm$.094 & 1.3$\pm$1.3   & 3.7$\pm$3.0   & 2.8$\pm$3.4  & 5.0$\pm$4.9  \\
                      & SAC                 &                 & \checkmark    &                  & \checkmark      & .444$\pm$.100 & 2.8$\pm$2.6   & 4.0$\pm$3.5   & 2.9$\pm$3.7  & 4.5$\pm$4.0  \\ \cline{2-11} 
                      & PPO                 & \checkmark      &               & \checkmark       &                 & .420$\pm$.115 & 7.2$\pm$10.6  & 4.4$\pm$3.4   & 3.0$\pm$2.7  & 5.3$\pm$3.6  \\
                      & PPO                 &                 &\checkmark     & \checkmark       &                 & .438$\pm$.103 & 4.1$\pm$5.9   & 4.7$\pm$4.7   & 2.7$\pm$1.7  & 7.1$\pm$6.1  \\
                      & PPO                 &                 & \checkmark    &                  & \checkmark      & .443$\pm$.097 & 1.8$\pm$1.0   & 3.9$\pm$3.5  & 2.8$\pm$1.9  & 5.9$\pm$4.7 \\\hline
\end{tabular}
\end{table}

\subsection{Comparing Results and Discussion}

 As illustrated in Table~\ref{tab:vs}, the strong negative correlation between the reward and the MAE of rotation and translation proves the efficacy of the overlap calculation reward function. Compared to the pre-training mode, the registration performance of online learning is superior. Online learning allows the model to continually refine its policy and update the highest reward. The registration result is determined by selecting the spatial transformation parameters that achieve the highest reward value across all iterations. However, online learning requires a significant amount of time. Table~\ref{tab:timevs} presents the hyperparameter experiments on timesteps for \textit{OL+SAC+Fuse+PCM}. Optimal performance was observed when timesteps were set at 10,000, yet the duration for online learning extends to 378 seconds. In comparison, loading a pre-trained model for direct registration takes only 10 seconds.

\begin{table}[tbp]
\centering
\caption{ Results of the timesteps hyperparameter experiment}\label{tab:timevs}
\renewcommand\arraystretch{1.3}
\tabcolsep3pt
\scriptsize
\begin{tabular}{ccccccc}
\hline
\multirow{2}{*}{Timesteps} & \multirow{2}{*}{Reward $\uparrow$} & \multicolumn{4}{c}{MAE $\downarrow$}                           & \multirow{2}{*}{Time(s)} \\ \cline{3-6}
                              &                          & $T_y$ (mm)     & $T_x$ (mm)  & $R_z$  ($^\circ$)     & $R_y$ ($^\circ$)              &                          \\ \hline
1000                          & .342$\pm$.122              & 27$\pm$35.1  & 9.6$\pm$11.1 & 4.9$\pm$4.15 & 6.6$\pm$5.1          & 37                       \\
2000                          & .393$\pm$.107              & 11.0$\pm$16.4  & 7.4$\pm$5.9  & 1.4$\pm$3.9  & 6.4$\pm$5.5          & 73                       \\
5000                          & .420$\pm$.103              & 6.7$\pm$12.4 & 4.8$\pm$4.4  & 3.4$\pm$3.7  & 6.2$\pm$4.5          & 193                      \\
10000                         & .444$\pm$.100              & 2.8$\pm$ 2.6  & 4.0$\pm$3.5  & 2.9$\pm$3.7  & 4.5$\pm$4.0 & 378                      \\
20000                         & .442$\pm$.100              & 3.6$\pm$6.9  & 4.9$\pm$4.0  & 3.3$\pm$2.8  & 5.9$\pm$7.9          & 750                      \\ \hline
\end{tabular}
\end{table}

\begin{figure}[t]
\centering
 \includegraphics[scale=0.42]{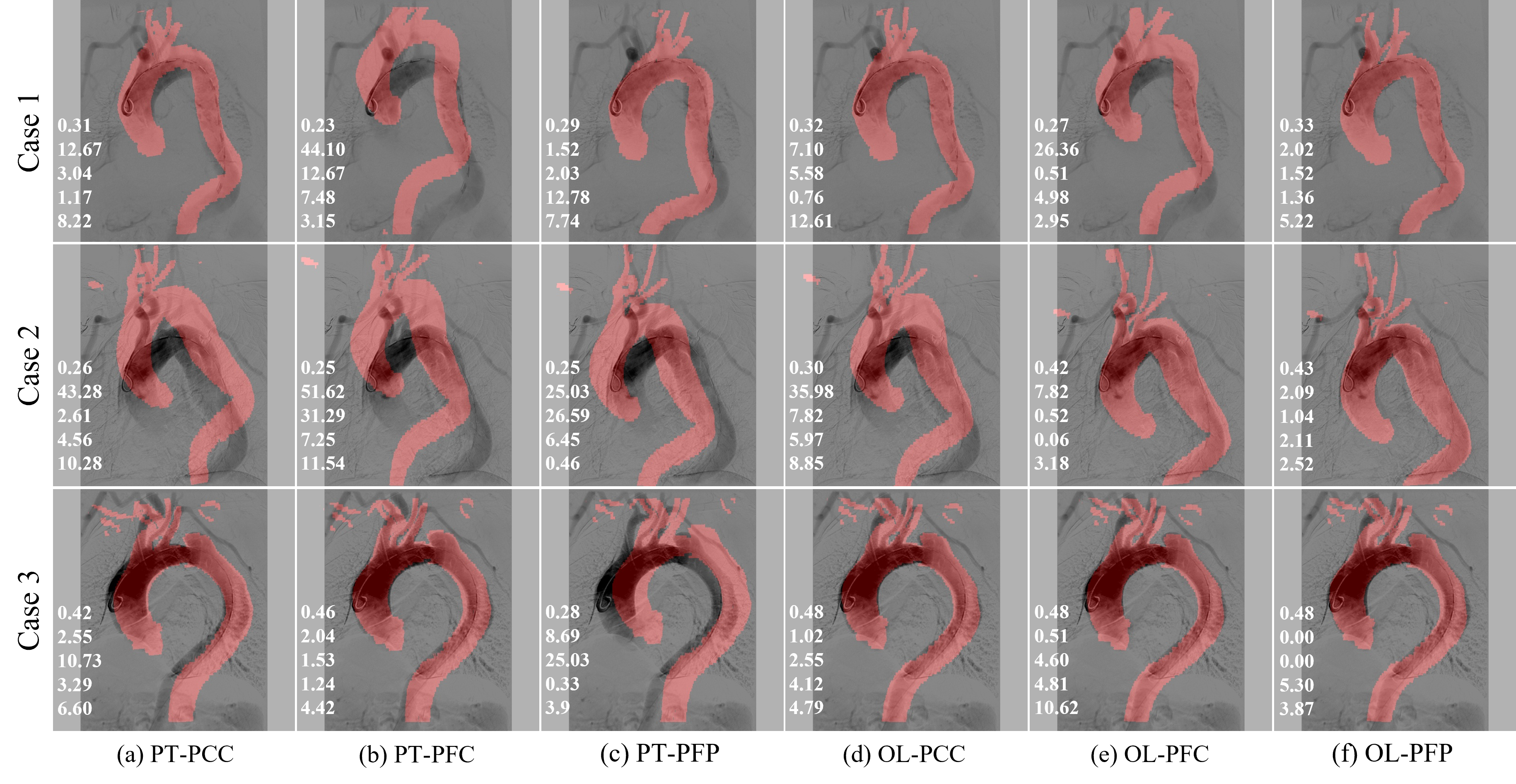}
\caption{Visualization of registration results using the PPO algorithm. (PCC: \textit{PPO+Concat+CNN}, PFC: \textit{PPO+Fuse+CNN}, PFP: \textit{PPO+Fuse+PCM})} \label{fig:show}
\end{figure}

Leveraging the encouragement of exploration through maximum entropy, The registration performance of SAC in online learning surpasses that of PPO. Meanwhile, PPO demonstrates more robust generalization capabilities, consistently outperforming SAC across various metrics in pre-training model. It is noteworthy that within the same mode and using the same RL algorithm, 
the configuration of \textit{Fuse+PCM} achieve the optimal results. We have visualized the results of the PPO experiments, as shown in Fig.~\ref{fig:show}. The first two rows correspond to cases of aortic dissection, while the last row pertains to a case of aortic aneurysm. The final column showcases the registration results under the online learning mode of \textit{PPO+Fuse+PCM}, where the red denotes the binary vascular map obtained through spatial transformation and parallel projection of the 3D CTA segmentation volume, essentially achieving comprehensive overlay of the DSA angiography area. The MAE for rotation and translation is relatively small. In comparison, other methods exhibit suboptimal registration performance in certain cases, particularly with larger absolute errors in $T_y$.

% \subsection{Discussion}

Despite the presence of artifacts and incomplete visualization of the aorta in DSA, the experimental results demonstrate that UDCR achieves accurate aortic DSA/CTA registration. In this work, we did not develop a new RL algorithm. Instead, we constructed a cross-dimensional registration environment based on the imaging principles and characteristics of DSA and CTA. It is noteworthy that this method is independent of ground truth or synthetic data and does not necessitate segmentation or centerline extraction of DSA. Only the segmentation of CTA requires additional annotated data. Furthermore, the advantage of UDCR based on preoperative 3D image segmentation is that it is not limited to data modalities, 3D rotational DSA and magnetic resonance angiography, along with other 3D imaging modalities, can also be registered with 2D DSA using UDCR. Furthermore, the results of the experiments is based on calculations performed with the current computing equipment and does not incorporate additional data acceleration or parallel processing techniques. Therefore, there is significant potential for improvement to reduce the algorithm's runtime.

\section{Conclusions}

In this work, we propose an unsupervised aortic DSA/CTA rigid registration method using DRL, which is independent of ground truth or synthetic data and the extraction of landmark of DSA. Specifically, we have established a cross-dimensional 2D/3D DRL registration environment based on spatial transformations, predicated on the imaging principles of DSA. Importantly, we introduce a novel overlap degree calculation reward function to evaluate the registration accuracy between segmentation maps and DSA images. This method is highly flexible, allowing for the loading of pre-trained models to perform registration directly or to seek the optimal spatial transformation parameters through online learning. Experiments on clinical DSA/CTA of the aorta indicate that the proposed UDCR achieved a MAE of 2.85mm in translation and a MAE of 4.35° in rotation, shows promising potential for application in intraoperative guidance. 

% \subsubsection{Acknowledgment.}This work was supported in part by the National Key R\&D Program of China (No.2018AAA0102600) and National Natural Science Foundation of China (No.62002082).

% the environments 'definition', 'lemma', 'proposition', 'corollary',
% 'remark', and 'example' are defined in the LLNCS documentclass as well.
%

% For citations of references, we prefer the use of square brackets
% and consecutive numbers. Citations using labels or the author/year
% convention are also acceptable. The following bibliography provides
% a sample reference list with entries for journal
% articles~\cite{ref_article1}, an LNCS chapter~\cite{ref_lncs1}, a
% book~\cite{ref_book1}, proceedings without editors~\cite{ref_proc1},
% and a homepage~\cite{ref_url1}. Multiple citations are grouped
% \cite{ref_article1,ref_lncs1,ref_book1},
% \cite{ref_article1,ref_book1,ref_proc1,ref_url1}.
% \bibliographystyle{unsrt}
\bibliographystyle{splncs04}
\bibliography{refer}

\begin{thebibliography}{10}
\providecommand{\url}[1]{\texttt{#1}}
\providecommand{\urlprefix}{URL }
\providecommand{\doi}[1]{https://doi.org/#1}

\bibitem{guan2020transfer}
Guan, S., Wang, T., Sun, K., Meng, C.: Transfer learning for nonrigid 2d/3d cardiovascular images registration. IEEE Journal of Biomedical and Health Informatics  \textbf{25}(9),  3300--3309 (2020)

\bibitem{huang2024real}
Huang, D.X., Zhou, X.H., Xie, X.L., Liu, S.Q., Feng, Z.Q., Hou, Z.G., Ma, N., Yan, L.: Real-time 2d/3d registration via cnn regression and centroid alignment. IEEE Transactions on Automation Science and Engineering  (2024)

\bibitem{nnunet}
Isensee, F., Jaeger, P.F., Kohl, S.A., Petersen, J., Maier-Hein, K.H.: nnu-net: a self-configuring method for deep learning-based biomedical image segmentation. Nature methods  \textbf{18}(2),  203--211 (2021)

\bibitem{jaganathan2023self}
Jaganathan, S., Kukla, M., Wang, J., Shetty, K., Maier, A.: Self-supervised 2d/3d registration for x-ray to ct image fusion. In: Proceedings of the IEEE/CVF Winter Conference on Applications of Computer Vision. pp. 2788--2798 (2023)

\bibitem{dias}
Liu, W., Tian, T., Wang, L., Xu, W., Li, H., Zhao, W., Pan, X., Yang, H., Gao, F., Deng, Y., et~al.: Dias: A comprehensive benchmark for dsa-sequence intracranial artery segmentation. arXiv preprint arXiv:2306.12153  (2023)

\bibitem{phtrans}
Liu, W., Tian, T., Xu, W., Yang, H., Pan, X., Yan, S., Wang, L.: Phtrans: Parallelly aggregating global and local representations for medical image segmentation. In: International Conference on Medical Image Computing and Computer-Assisted Intervention. pp. 235--244. Springer (2022)

\bibitem{markelj2012review}
Markelj, P., Toma{\v{z}}evi{\v{c}}, D., Likar, B., Pernu{\v{s}}, F.: A review of 3d/2d registration methods for image-guided interventions. Medical image analysis  \textbf{16}(3),  642--661 (2012)

\bibitem{global1}
Mensah, G.A., Fuster, V., Murray, C.J., Roth, G.A.: Global burden of cardiovascular diseases and risks, 1990-2022. Journal of the American College of Cardiology  \textbf{82}(25),  2350--2473 (2023)

\bibitem{review}
Mitrovi{\'c}, U., Likar, B., Pernu{\v{s}}, F., {\v{S}}piclin, {\v{Z}}.: 3d--2d registration in endovascular image-guided surgery: Evaluation of state-of-the-art methods on cerebral angiograms. International journal of computer assisted radiology and surgery  \textbf{13},  193--202 (2018)

\bibitem{avt}
Radl, L., Jin, Y., Pepe, A., Li, J., Gsaxner, C., Zhao, F.h., Egger, J.: Avt: Multicenter aortic vessel tree cta dataset collection with ground truth segmentation masks. Data in brief  \textbf{40},  107801 (2022)

\bibitem{stable-baselines3}
Raffin, A., Hill, A., Gleave, A., Kanervisto, A., Ernestus, M., Dormann, N.: Stable-baselines3: Reliable reinforcement learning implementations. Journal of Machine Learning Research  \textbf{22}(268), ~1--8 (2021), \url{http://jmlr.org/papers/v22/20-1364.html}

\bibitem{su2022cave}
Su, R., van~der Sluijs, M., Chen, Y., Cornelissen, S., van~den Broek, R., van Zwam, W., van~der Lugt, A., Niessen, W., Ruijters, D., van Walsum, T.: Cave: Cerebral artery-vein segmentation in digital subtraction angiography. arXiv preprint arXiv:2208.02355  (2022)

\bibitem{svecic2021multimodal}
Svecic, A., Soulez, G., Monet, F., Kashyap, R., Kadoury, S.: Multimodal sensing guidewire for c-arm navigation with random uv enhanced optical sensors using spatio-temporal networks. In: Medical Image Computing and Computer Assisted Intervention--MICCAI 2021: 24th International Conference, Strasbourg, France, September 27--October 1, 2021, Proceedings, Part IV 24. pp. 249--258. Springer (2021)

\bibitem{CAR-Net}
Wu, W., Zhang, J., Peng, W., Xie, H., Zhang, S., Gu, L.: Car-net: a deep learning-based deformation model for 3d/2d coronary artery registration. IEEE Transactions on Medical Imaging  \textbf{41}(10),  2715--2727 (2022)

\bibitem{zhu2021iterative}
Zhu, J., Li, H., Ai, D., Yang, Q., Fan, J., Huang, Y., Song, H., Han, Y., Yang, J.: Iterative closest graph matching for non-rigid 3d/2d coronary arteries registration. Computer Methods and Programs in Biomedicine  \textbf{199},  105901 (2021)

\bibitem{zhu2019monte}
Zhu, J., Song, S., Guo, S., Ai, D., Fan, J., Song, H., Liang, P., Yang, J.: Monte carlo tree search for 3d/2d registration of vessel graphs. In: 2019 IEEE International Conference on Bioinformatics and Biomedicine (BIBM). pp. 787--791. IEEE (2019)

\end{thebibliography}
% \end{thebibliography}

\end{document}